\documentclass[a4paper,12pt]{article}

\usepackage{ragged2e} 
\usepackage{changepage}

\usepackage{amsfonts}
\newcommand{\N}{\mathbb{N}}
\usepackage{siunitx} 
\usepackage{amsmath}
\usepackage{bm} 
\usepackage{amssymb}
\usepackage{wasysym}
\usepackage{xfrac} 

\usepackage{graphicx}
\usepackage{tikz}
\usetikzlibrary{positioning, calc}
\usepackage{sidecap}
\usepackage{wrapfig}
\usepackage{caption}
\usepackage{subcaption}

\usepackage{verbatim}
\usepackage[english]{babel}

\usepackage{array}
\usepackage{adjustbox}
\usepackage{float}
\usepackage{enumerate}
\usepackage{multirow}
\usepackage{multicol}
\usepackage[table,xcdraw]{xcolor}

\usepackage[utf8]{inputenc}
\usepackage[super,sort&compress]{natbib}
\makeatletter
\renewcommand{\@biblabel}[1]{#1.}
\makeatother
\setcitestyle{comma}
\usepackage{csvsimple}
\usepackage{bbm}
\usepackage{pdflscape}
\usepackage{breqn}
\usepackage{lscape}
\usepackage{amsfonts}
\usepackage{latexsym}
 \usepackage[titletoc]{appendix}  
\usepackage[colorlinks=true, citecolor= black]{hyperref} 
\hypersetup{linkcolor=black} 
\usepackage[all]{hypcap} 
\usepackage{setspace}
\usepackage{enumitem}
\usepackage{authblk}
\usepackage{bm}
\usepackage[T1]{fontenc}
\usepackage{helvet}
\usepackage{listings}
\lstdefinestyle{Rstyle}{
  language=R,
  basicstyle=\ttfamily\scriptsize,
  numbers=left,
  numberstyle=\tiny,
  numbersep=5pt,
  breaklines=true,
  breakatwhitespace=true,
  tabsize=2,
  frame=single,
  showstringspaces=false
}
\usepackage[pagewise]{lineno}

\title{\textbf{Federated generalized linear mixed models based on one-time shared summary statistics}}
\author[1]{\small Marie Analiz April Limpoco}
\author[1]{\small Christel Faes}
\author[1,2]{\small Niel Hens}
\affil[1]{Interuniversity Institute for Biostatistics and statistical Bioinformatics (I-BioStat), Data Science Institute (DSI), Hasselt University, Hasselt, Belgium}
\affil[2]{Centre for Health Economic Research and Modelling Infectious Diseases (CHERMID), Vaccine \& Infectious Disease Institute, Antwerp University, Antwerp, Belgium}

\date{}
\newtheorem{theorem}{Theorem}[section]
\begin{document}
\maketitle


\begin{abstract}
Data privacy has increasingly become a daunting challenge because it limits data availability, which is essential in estimating statistical models such as generalized linear mixed models. Access to personal data often involves considerable time, effort, and paperwork, which can impede research progress and collaboration. Existing approaches that do not use individual-level data for model estimation are either prone to ecological bias, cannot handle heterogeneity, or require iterative communication. In this paper, we propose an approach to estimate generalized linear mixed models based on summary statistics shared only once. We used linear, logistic, and Poisson mixed models as examples to demonstrate the methodology. Our strategy involves generating pseudo-data whose summary statistics match those of the actual but unavailable data. These pseudo-data are then used for model estimation instead of the actual data. The estimates we achieve are identical (up to the third decimal place) to those derived from actual data and have similar bias, coverage, and prediction performance. Communication and resource efficiency distinguish our approach from existing methods.
\end{abstract}


{\small \bf Correspondence:} {\small Marie Analiz April Limpoco \\ 
Interuniversity Institute for Biostatistics and statistical Bioinformatics (I-BioStat),\\ Data Science Institute (DSI), Hasselt University, Hasselt, Belgium \\
Email: \textcolor{blue}{liz.limpoco@uhasselt.be}}

\section{Introduction}
Generalized linear mixed models (GLMM) are invaluable tools for health research involving data from multiple data providers. They address the heterogeneity introduced by differences between data providers. These models offer flexibility in analyzing health data composed of different types of variables. For example, analyzing patient records from different hospitals may involve identifying risk factors that affect tumor size. Since tumor size is a continuous variable, linear mixed models can be fitted to the data. Researchers may also be interested in the probability of cancer incidence, which can be predicted using logistic mixed models. Modeling the number of hospital admissions may help hospitals plan and allocate resources, in which case Poisson mixed models can be used. These models offer essential knowledge to health professionals. However, estimation of these models requires individual-level data. \\

With the increasing threat and far-reaching repercussions of sensitive data disclosure especially in health research \cite{healthcare8020133}, data acquisition has become more demanding than in the past. Privacy regulation has become more stringent, especially for requests concerning personal health data. As a result, health research may be delayed and reproducibility of results may take more time to establish. This privacy constraint underscores the need to develop strategies to estimate statistical models without accessing individual-level data; that is, while data remain federated. \\

Aggregate data (AD) meta-analysis offers strategies to handle this challenge. It enables inference about the parameters of interest based on summary information alone, thus preserving the confidentiality of individual observations. Despite its benefits, AD meta-analysis also has limitations. In particular, inference from AD meta-analysis is known to be prone to ecological bias due to the use of group-level instead of individual-level characteristics in the analysis \citep{https://doi.org/10.1002/sim.1187, https://doi.org/10.1002/sim.1023}. Moreover, normality is typically assumed within each study and within-study variances are assumed to be known. These assumptions can be problematic for small studies \citep{Burke2017855}. When results are compared to a one-stage individual participant data (IPD) meta-analysis, the gold standard in meta-analysis, inference from AD meta-analysis may \citep{Tudur_Smith2016-nw} or may not agree \citep{10.1371/journal.pone.0060650}, which may lead to confusion. \\

In the literature, there are other methods to estimate a GLMM that produce individual-level inference without accessing individual-level data, but they also have their respective drawbacks. Some of them involve more than one communication round between data providers and data analysts before converging to a global model \citep{10.1093/bioinformatics/btaa478, 10.1093/jamia/ocac067, Li2022, 10.1371/journal.pone.0280192}. Other methods utilize aggregated data shared only once and draw conclusions at the individual-level, but cannot handle heterogeneity \citep{doi:10.1142/9789813279827_0004, 10.1093/jamia/ocaa044} or only work under Gaussian assumptions \citep{f6c8409ac3ed4b1cb27df3dfa95ac12c, https://doi.org/10.1002/sim.10300}. In this paper, we fill the gap in the literature by proposing a novel strategy that overcomes the limitations of the aforementioned methods.  \\

Our proposed strategy builds on the strengths of the existing methods in the literature. It supports individual-level analysis while requiring summary statistics shared only once. It takes into account heterogeneity in data providers by including random effects in the models. We show that this approach is generalizable to members of the exponential family of distributions, and we demonstrate it using linear, logistic, and Poisson mixed models as concrete examples. In a nutshell, our proposed strategy enables GLMM estimation without accessing actual individual-level data from data providers. \\        
    
To provide details on how the proposed strategy works and how well it performs, the subsequent sections are organized as follows. We first develop the methodology in Section \ref{section:methods}. We then assess the performance of our proposed strategy relative to the classical method that requires pooled individual observations (Section \ref{section:simres}). Following this assessment, a practical implementation is demonstrated in Section \ref{section:realexample} through real, publicly available, de-identified data. These data are hospital inpatient discharge data collected by the Statewide Planning and Research Cooperative System (SPARCS) in 2022 \href{https://health.data.ny.gov/Health/Hospital-Inpatient-Discharges-SPARCS-De-Identified/5dtw-tffi/about_data}{(\url{https://health.data.ny.gov/})} and published by the New York State Department of Health online. All R scripts and necessary material are available in the Github repository found in this link: \href{https://github.com/lizlimpocouhasselt/FederatedGLMM/}{\url{https://github.com/lizlimpocouhasselt/FederatedGLMM}}. Finally, Section \ref{section:discussion} discusses the results and provides implications and insights.

\section{Theory and methods}\label{section:methods}

\begin{figure}[h]
    \centering
    \begin{tikzpicture}[
        node distance=0.75cm, 
        block/.style={
        rectangle,
        draw,
        text width=10em,
        text centered,
        rounded corners,
        minimum height=2em
        },
        arrow/.style={
        thick,
        ->,
        >=stealth,
        draw=black!70
        }
    ]

    \node (data1) [block] {Individual-level data 1};
    \node (data2) [block, right= of data1] {Individual-level data 2};
    \node (data3) [block, right= of data2] {Individual-level data 3};

    \node (summary1) [block, below= of data1] {Summary statistics 1};
    \node (summary2) [block, right= of summary1] {Summary statistics 2};
    \node (summary3) [block, right= of summary2] {Summary statistics 3};

    \node (pseudo1) [block, below= of summary1] {Pseudo-data 1}; 
    \node (pseudo2) [block, right =of pseudo1] {Pseudo-data 2}; 
    \node (pseudo3) [block, right =of pseudo2] {Pseudo-data 3}; 

    \node (estimation) [block, below = of pseudo2] {Model estimation};

    \draw [arrow] (data1) -- (summary1);
    \draw [arrow] (data2) -- (summary2);
    \draw [arrow] (data3) -- (summary3);

    \draw [arrow] (summary1) -- (pseudo1);
    \draw [arrow] (summary2) -- (pseudo2);
    \draw [arrow] (summary3) -- (pseudo3);

    \draw [arrow] (pseudo1) -- (estimation);
    \draw [arrow] (pseudo2) -- (estimation);
    \draw [arrow] (pseudo3) -- (estimation);

    \node[rotate=90, anchor=south] at ($(summary1.west)+(-0.75cm, -0.74cm)$) {\textbf{Data analyst | Data provider}};

    \draw[dashed, thick] 
    ($(summary1.south west) + (-0.5cm, -0.4cm)$) -- 
    ($(summary1.south west)!1!(summary3.south east)+ (-0.5cm, -0.4cm)$);
    \end{tikzpicture}
    \caption{Proposed framework for a setup with three data providers. Each data provider aggregates data into summary statistics. The data analyst generates pseudo-data whose summary statistics match those supplied by the data providers. A global model can then be estimated from the pooled pseudo-data.}
    \label{fig:methoddiag}
\end{figure}
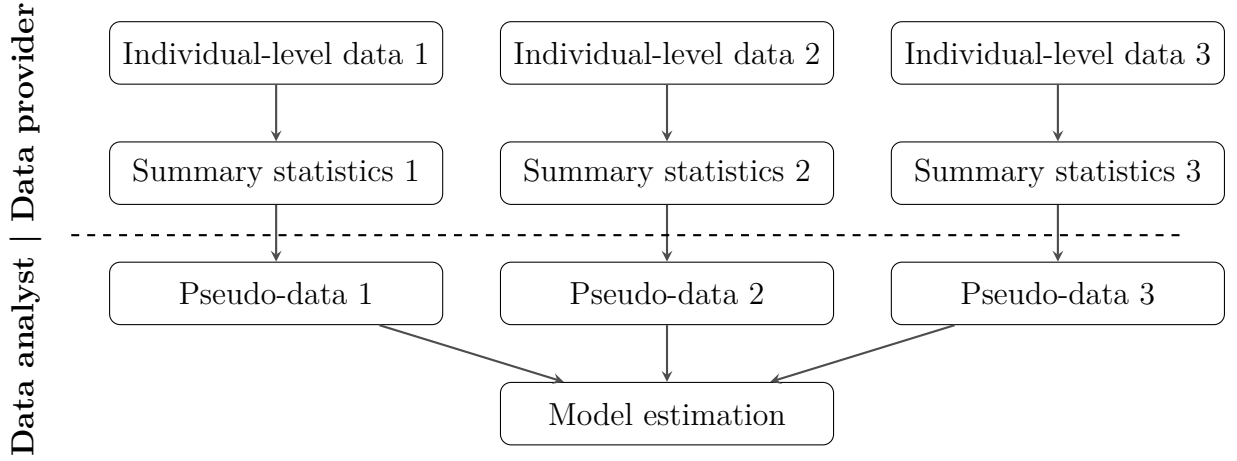

In this section, we develop a framework (Figure \ref{fig:methoddiag}) to estimate generalized linear models (GLM), with or without random effects, from summary statistics shared only once by one or more data providers. We explain which summary statistics are informative enough to infer about the parameters of interest. From these summary statistics computed by each data provider on the complete individual-level data, we propose to generate pseudo-data. Pseudo-data generation will be performed by the data analyst, who will use pseudo-data for model estimation. Lastly, we provide details of the simulations we conducted to evaluate the performance of our proposed framework relative to the classical method, which uses pooled individual-level observations.

\subsection{Polynomial-approximate sufficient statistics (PASS) of the exponential family}\label{suf}
To specify which summary statistics data providers must convey to enable data analysts to infer about the parameters of interest, we begin with the principle of statistical sufficiency. This principle allows data reduction without losing important information about the parameters of interest \cite{CaseBerg:01}. It permits parameter estimation using sufficient statistics only instead of requiring individual-level data. However, not all parametric models have sufficient statistics. \\

For models with no sufficient statistics and whose assumed distribution comes from the exponential family, we propose to use \textit{polynomial-approximate} sufficient statistics (PASS). In Limpoco, et. al. (2025)\citep{https://doi.org/10.1002/bimj.70080}, PASS were defined as summary statistics that can be used to construct the polynomial expression of the log-likelihood of a logistic mixed model. In this paper, we generalize PASS to the exponential family of distributions. \\

Generalizing to the exponential family requires examining the associated log-likelihood. We write the log-likelihood of any member of the exponential family as

\begin{align}\label{eq:expfam_loglik}
        l(\boldsymbol{\theta}, \phi) &= \frac{1}{a(\phi)}\left[ \sum_{i=1}^n y_i\theta_i - \sum_{i=1}^n b(\theta_i)\right] + \sum_{i=1}^n c(y_i, \phi),
\end{align}

where $y_i$ is the $i$th data point ($i = 1,...,n$); $\boldsymbol{\theta}$ is the vector of parameters related to the mean; $\phi$ is the dispersion parameter; and $a(.)$, $b(.)$, and $c(.)$ are known functions that further characterize each member of the family. Assuming availability of individual-level data and a specified regression model, this log-likelihood is constructed using empirical data consisting of the response vector $\mathbf{y}$ and the design matrix $\mathbf{X}$. These empirical data enter the log-likelihood through $\sum_{i=1}^n y_i\theta_i$, $\sum_{i=1}^n b(\theta_i)$, and $\sum_{i=1}^n c(y_i, \phi)$. Examining these summations further will reveal which summary statistics can be used to replace individual-level data in the model estimation process. To be more concrete, we study the case for Gaussian, Bernoulli, and Poisson distributions (Table \ref{tab:expdist}) in the context of regression models where predictor data $\mathbf{X}$ are assumed to be fixed.

\begin{table}[h]
    \centering
    \caption{Characteristics of three members of the exponential family when used in regression models}
    \begin{tabular}{lcccc}
        \hline
        Distribution & $\theta_i$ & $b(\theta_i)$ & $\phi$ & $c(y_i, \phi)$ \\
        \hline
        &&&&\\
        Gaussian & $\mathbf{x}_i^T \boldsymbol{\beta}$ & $\frac{(\mathbf{x}_i^T \boldsymbol{\beta})^2}{2}$ & $\sigma^2$ & $\frac{-(y_i^2/\sigma^2 + \log{(2\pi\sigma^2)})}{2}$ \\
        &&&&\\
        Bernoulli & $\mathbf{x}_i^T \boldsymbol{\beta}$ & $\log{(1 + \exp{(\mathbf{x}_i^T \boldsymbol{\beta})})}$ & 1 & 0 \\
        &&&&\\
        Poisson & $\mathbf{x}_i^T \boldsymbol{\beta}$ & $\exp{(\mathbf{x}_i^T \boldsymbol{\beta})}$ & 1 & $-\log{(y_i!)}$ \\
        \hline
    \end{tabular}
    \label{tab:expdist}
\end{table}

Given the characteristics of models defined under the Gaussian, Bernoulli, and Poisson assumptions, we can identify summary statistics that can be used to construct the first ($\sum_{i=1}^n y_i\theta_i$) and last ($\sum_{i=1}^n c(y_i, \phi)$) summations in the log-likelihood. In particular, these summary statistics are the sample mean ($\bar{\mathbf{y}}$) and variance ($s^2_{\mathbf{y}}$) of the response vector $\mathbf{y}$; sample covariances ($s_{\mathbf{yx}_j}$) between $\mathbf{y}$ and each of the $p$ predictors $\mathbf{x}_j$, $j=1,...,p$; and $\sum{\log{(\mathbf{y}!)}}$. However, this task of identifying summary statistics for the second summation ($\sum_{i=1}^n b(\theta_i)$) is not as easy. \\

The case for Gaussian models is more explicit than for Bernoulli and Poisson models. Limpoco et. al. \citep{https://doi.org/10.1002/sim.10300} have shown that $\sum_{i=1}^n b(\theta_i)$ can be derived from the covariance matrix computed on the predictor data $\mathbf{X}$ for Gaussian models. For Bernoulli and Poisson models, the summation involves logarithmic and exponential functions, which do not allow the summation to directly operate on the empirical data. Thus, we need to rewrite $\sum_{i=1}^n b(\theta_i)$ in another form where the summary statistics are more explicit. \\

Rewriting the second summation reveals that, in general, the sample moments (Appendix \ref{app:sample_moments}) comprise the log-likelihood of the Gaussian, Bernoulli, and Poisson models. Using the Weierstrass approximation theorem (Appendix \ref{app:theorems}), $\sum_{i=1}^n b(\theta_i)$ becomes
\begin{align}\label{eqn:2ndsum}
    \sum_{i=1}^n b(\mathbf{x}_i^T \boldsymbol{\beta}) &= \sum_{k=0}^K{\alpha_k \sum_{i=1}^n(\mathbf{x}_i^T \boldsymbol{\beta})^k} + \sum_{i=1}^n \epsilon_i,
\end{align}
where $K$ is the polynomial degree, $\alpha_k$ is the coefficient associated with the $k$th term of the polynomial, and $\epsilon_i$ is the polynomial approximation error. The current form of $\sum_{i=1}^n b(\theta_i)$ explicitly shows that it can be computed from sample moments. Since the summary statistics for the other two summations in Equation \ref{eq:expfam_loglik} are also sample moments, we claim that the full log-likelihood of Gaussian, Bernoulli, and Poisson models is made up of sample moments. Thus, we say that knowing the sample moments is sufficient for model estimation. However, this sufficiency of sample moments is exact for Gaussian models only. For Bernoulli and Poisson models, the sample moments are only sufficient for the polynomial part of the re-expressed log-likelihood. \\

To indicate the limitation of the sufficiency of sample moments, we use the term polynomial-approximate sufficient statistics (PASS) to refer to these summary statistics. As such, we consider sample moments as exact sufficient statistics for Gaussian models but PASS for Bernoulli and Poisson models. In general, we have identified that sample moments are the summary statistics that must be shared between data providers and data analysts to enable model estimation without individual-level data. For a more technical derivation, see Appendix \ref{app:poly_loglik}.

\subsection{PASS of generalized linear mixed models (GLMM)}
We extend the same reasoning in the previous section to derive summary statistics that are PASS for models handling grouped data. For any generalized linear model with random effects $\mathbf{u}_i$ ($i = 1,..., m$ where $m$ is the number of groups), the conditional probability of the response vector ($\mathbf{Y}_i$) per group $i$ with size $n_i$ at fixed values of design matrix $\mathbf{X}_i$ is
\begin{align*}
    g(\mathbf{Y}_i|\mathbf{X}_i,\mathbf{u}_i;\boldsymbol{\beta}, \phi) &= \prod_{j=1}^{n_i} p(Y_{ij}|\mathbf{x}_{ij},\mathbf{u}_i;\boldsymbol{\beta}, \phi),
\end{align*}
where $\boldsymbol{\beta}$ and $\phi$ are the parameters of the specified distribution, $p(\cdot)$ is the probability density function of $Y_{ij}$, and $\mathbf{x}_{ij}$ is the $j$th row in $\mathbf{X}_i$. Given the response data $\mathbf{y}_i$ and random effects design matrix $\mathbf{Z}_i$, the marginal likelihood function of the parameters is
\begin{align*}
    L_i(\boldsymbol{\beta}, \phi, \boldsymbol{\zeta}|\mathbf{y}_i,\mathbf{X}_i, \mathbf{Z}_i) &= \int{g(\mathbf{Y}_i|\mathbf{X}_i,\mathbf{u}_i;\boldsymbol{\beta}, \phi)f(\mathbf{u}_i;\boldsymbol{\zeta})d\mathbf{u}_i},
\end{align*}
where $f(\mathbf{u}_i;\boldsymbol{\zeta})$ is the probability distribution of $\mathbf{u}_i$ parameterized by $\boldsymbol{\zeta}$. Here, individual-level data enter the log-likelihood through $g(\cdot)$, which can be expressed as a polynomial up to a certain level of accuracy $\epsilon$. For example, in a Poisson mixed model where $\phi = 1$,
\begin{align*}
        g(\mathbf{Y}_i|\mathbf{X}_i,\mathbf{u}_i;\boldsymbol{\beta}) &= \exp{\left(\sum_{j=1}^{n_i} y_{ij}\eta_{ij} - \sum_{j=1}^{n_i} \exp{(\eta_{ij})} - \sum_{j=1}^{n_i}\log{(y_i!)}\right)},
    \end{align*}
where $\eta_{ij} = \mathbf{x}_{ij}^T\boldsymbol{\beta} + \mathbf{z}_{ij}^T\mathbf{u}_i$. With respect to $\eta_{ij}$, the first and third summations ($\sum_{j=1}^{n_i} y_{ij}\eta_{ij}$ and $\sum_{j=1}^{n_i}\log{(y_i!)}$) are already polynomials with degree 1 and 0, respectively. Using Weierstrass approximation theorem (Appendix \ref{app:theorems}), the second term $\sum_{j=1}^{n_i} \exp{(\eta_{ij})}$ can also be expressed as a polynomial up to an error, as in Section \ref{suf}. It follows that the full likelihood
\begin{align*}
    L &= \prod_{i=1}^m L_i(\boldsymbol{\beta}, \phi, \boldsymbol{\zeta}|\mathbf{y},\mathbf{X}),
\end{align*}
can also be expressed as a polynomial up to a certain level of accuracy. Consequently, $L$ is comprised of sample moments (Appendix \ref{app:poly_loglik}). Thus, sample moments are also PASS for generalized linear mixed models. This result implies that sample moments computed per group contain the necessary information to enable parameter estimation.

\subsection{Pseudo-data generation}\label{section:psgen}
Given the sample moments of the actual but unavailable data, we propose to generate what we refer to as \textit{pseudo-data} instead of directly using sample moments for model estimation. These pseudo-data are artificially generated individual-level data whose sample moments match those of the actual data. When generating pseudo-data, we do not aim to reconstruct the actual data; rather, we aim to facilitate model estimation using an individual-level form of the information contained in sample moments. By doing so, we exploit the existing model estimation functionalities in statistical software programs which expect individual-level observations. More importantly, pseudo-data generation allows fitting a wider range of statistical models. Specifically, once pseudo-data are generated, statistical models assuming distributions from the exponential family can be fitted, thus making our strategy generalizable. \\

To achieve flexibility in the statistical models that can be fit, we propose to generate pseudo-data for all variables without constraining any of them to a particular data type. In contrast to the paper of Limpoco et. al. (2025) \citep{https://doi.org/10.1002/bimj.70080}, our proposed strategy does not need to constrain pseudo-responses. In particular, we relax binary or integer constraints for all variables during pseudo-data generation. Log-likelihood functions for Gaussian, Bernoulli, and Poisson models can still be constructed even when these constraints are not imposed. This relaxation results in a much closer match between the sample moments of the actual and pseudo-data than when constraints are imposed. Consequently, the log-likelihood constructed from pseudo-data approximates that of the actual data much better, leading to more similar performance in model estimation. Additionally, we do not need to specify the model to use before generating pseudo-data since there is no need to identify and constrain pseudo-responses. Thus, the generated pseudo-data can be used for any model estimation tasks such as fitting Gaussian, Bernoulli, or Poisson models. \\

To generate pseudo-data to fit any GLM, we employ unconstrained nonlinear least squares optimization. Specifically, we minimize the sum of squared differences (SSD) between sample moments of the actual data $\bar{\mu}^{(\mathbf{r})}_d$ and those of the pseudo-data $\bar{\mu}^{(\mathbf{r})}_{ps}$:
\begin{align*}
       \text{SSD} &= \sum_{\substack{\mathbf{r} \epsilon \N_0^p\\ 1\leq |\mathbf{r}| \leq K}} (\bar{\mu}^{(\mathbf{r})}_d - \bar{\mu}^{(\mathbf{r})}_{ps})^2.  
\end{align*}
Given $p$ variables in the actual data, we include sample moments of order $\mathbf{r} = (r_1,r_2,...,r_p)$ $\epsilon$ $\N_0^p$ from $|\mathbf{r}| = 1$ up to $K$. This $K$ value will be determined through simulations which will be described in Section~\ref{section:simdesc}. For $n$ actual observations, we generate $n$ pseudo-observations. In total, we have to generate $np$ pseudo-values. \\

The total number of pseudo-values to be generated ($np$) and the number of sample moments to be matched, which we denote here as $\nu$, dictate the nature of the solution obtained from the optimization process. For example, whenever $np > \nu$ where
\begin{align*}
    \nu &= 4p + \frac{p^2 - p}{2} + 5{p \choose 2} + 4{p \choose 3} + {p \choose 4}
\end{align*}
for $K=4$ and $p \geq 4$, the optimization problem becomes an underdetermined system. As a consequence, there will be infinitely many pseudo-data sets that can correspond to the same set of sample moments. This system guarantees that the probability of exactly recovering the actual dataset is very low, which is favorable in terms of preventing the disclosure of sensitive data.  To this end, we use the Levenberg-Marquardt optimization algorithm which can handle underdetermined systems. For practical implementation, we use the function \texttt{lsqnonlin} from the \texttt{R}\citep{Rpackage} package \texttt{pracma}\citep{Rpracma}.

\subsection{Model estimation in \texttt{R}}
To use the model estimation functionalities in \texttt{R} to fit GLM[M]s on unconstrained pseudo-data, modifications had to be introduced. In \texttt{R}, binary and nonnegative integer constraints are imposed for Bernoulli and Poisson models, respectively. These constraints cause errors when unconstrained pseudo-data are entered as input to \texttt{R} functionalities that estimate GLM[M]s (e.g., \texttt{glm} and \texttt{glmer}). Since the constraints are specified for each model, we adapt the existing \texttt{family} definitions for \texttt{binomial} and \texttt{poisson} in \texttt{R} to relax these constraints. Table \ref{tab:customfam} summarizes the proposed modifications in the initialization, deviance residual function, and AIC computation.

\begin{table}[h]
    \centering
    \caption{Family definition modifications}
    \label{tab:customfam}
    \resizebox{\textwidth}{!}{%
        \begin{tabular}{lll}
            \hline
            Part of code & \texttt{binomial} & \texttt{poisson}\\
            \hline
            Name & \texttt{soft\_binomial} & \texttt{soft\_poisson}\\
            Initialization & removed \texttt{is($0<y<1$)} & 
             removed \texttt{is($y \geq 0$)} \\
             &  & initial $\lambda_i$ from $(y_i+0.1)$ to $\max{(y_i, 0.1)}$ \\
            Deviance residuals & dropped $logLik_{saturated}$ & dropped $logLik_{saturated}$\\
            AIC & dropped $\log{\binom{1}{y}}$ & dropped $\log{(y!)}$\\
        \hline
        \end{tabular}
    }
\end{table}
To justify the changes made to the deviance residual function, we examined how this function affects parameter estimation. In \texttt{R}, maximum likelihood (ML) estimation for GLM[M]s is implemented by minimizing the deviance function instead of directly maximizing the log-likelihood. The deviance function is the difference in log-likelihood between the fitted and saturated models 
\begin{align*}
    deviance &= -2(logLik_{fitted} - logLik_{saturated}).
\end{align*}
Since $logLik_{saturated}$ involves taking logarithms of the response data (Appendix \ref{app:loglik_sat}), negative pseudo-responses will return errors. However, $logLik_{saturated}$ does not contain parameters, so it does not affect ML estimates. Thus, we propose to drop $logLik_{saturated}$. 

The effect of dropping $logLik_{saturated}$ from the deviance residual function is different for each distribution. For Bernoulli models, we show in Appendix \ref{app:loglik_sat} that dropping $logLik_{saturated}$ has a negligible impact on the deviance residual function. Even when $y_i$ is constrained to be binary, $logLik_{saturated}$ is almost zero. In contrast, the case for Poisson models is not as convenient. The contribution of $logLik_{saturated}$ to the deviance residual function is only negligible when $y_i=0$. When $y_i > 0$, the deviance residual is
\begin{align*}
    2\left[y_i \log{\left(\frac{y_i}{\hat{\lambda}_i}\right)} - (y_i - \hat{\lambda}_i) \right],
\end{align*}
where $\hat{\lambda}_i$ is the fitted value for observation $i$. Nevertheless, for ML estimation, this is not an issue. \\    

Similarly, modifying the AIC computation to allow the use of unconstrained pseudo-data should not be an issue as long as model selection results are preserved. The AIC computation returns an error because of the terms $\log{\binom{1}{y}}$ for Bernoulli models and $\sum \log{(y!)}$ for Poisson models. Specifically, these terms cannot be computed for negative non-integer pseudo-responses. However, dropping them from the AIC formula will not affect model comparison. For Bernoulli models, the AIC value stays the same because $\log{\binom{1}{y}}$ is zero even when $y$ is binary. This is not the case for Poisson models, as the AIC value is not preserved when dropping $\sum \log{(y!)}$. Nevertheless, model selection will be unaffected since this term is constant for all models estimated on the same pseudo-data.


\subsection{Simulation study for Poisson mixed models}\label{section:simdesc}
With the modified statistical functionalities in \texttt{R}, we now have the tools to investigate up to which value of $K$ should be considered to produce models as good as those produced by the classical method. Recall that $K$ is the polynomial degree in the polynomial expression of the log-likelihood. It also corresponds to the highest order $|\mathbf{r}|$ of sample moments that should be matched between the actual and pseudo-data. For a model under the Gaussian distribution assumption, matching up to $|\mathbf{r}| = 2$ (i.e. mean and covariance) is sufficient to exactly replicate the numerical estimates that would have been produced from the actual data \citep{https://doi.org/10.1002/sim.10300}. For logistic mixed models, Limpoco, et. al. (2025) \citep{https://doi.org/10.1002/bimj.70080} showed that matching sample moments up to at least $|\mathbf{r}| = 3$ produces estimates that perform as good as the estimates derived from the actual data in terms of bias and coverage. In this paper, we perform simulation studies for Poisson mixed models, similar to those conducted by Limpoco, et. al. (2025), to determine a suitable value for $K$. \\ 

The simulation setup was as follows. We used the preliminary data analysis results on SPARCS 2022 as basis for the simulation parameters. The true Poisson mixed model, which has a random intercept $u_i$ per group $i$, is defined as follows:\\
\begin{align*}
    Y_{ij} &\sim \mathrm{Pois}(\lambda_{ij}) \\
    X_1 &\sim \mathrm{N}(\mu = 3150.14, \sigma^2 = 710797.5) \\
    X_2 &\sim \mathrm{Bern}(\pi=0.56) \\
    \mathbf{X}_3 &\sim \mathrm{Multinom} (N; \pi_1=0.095, \pi_2=0.127, \pi_3=0.111, \pi_4=0.387, \pi_5=0.280) \\
    u_i &\sim \mathrm{N}(0, \sigma^2_u) \\
    \log{(\lambda_{ij})} &= \mathbf{x}_{ij}^T \boldsymbol{\beta} + u_i \\
    \sigma_u = 0.48, \beta_0 &= 2.29, \beta_1 = -0.30, \\
    \beta_2 = 0.09, \beta_{32} &= -0.96, \beta_{33} = -0.81, \\
    \beta_{34} = -0.81, \beta_{35} &= -0.79.
\end{align*} 
We simulated 500 datasets for each of three settings: $m=30$ \& $n=100$; $m=50$ \& $n=60$; and $m=100$ \& $n=30$, where $m$ is the number of groups and $n$ is the uniform group size. For each simulated dataset, pseudo-data were generated following the framework described in Section~\ref{section:psgen}. In particular, three types of pseudo-data were generated based on $K = 2, 3,$ and 4. From here onward, we denote simulated data as $(\mathbf{y}, \mathbf{X})_d$ while $(\mathbf{y}, \mathbf{X})_{ps_2}$, $(\mathbf{y}, \mathbf{X})_{ps_3}$, and $(\mathbf{y}, \mathbf{X})_{ps_4}$ denote pseudo-data whose sample moments match those of the simulated data up to order $|\mathbf{r}| = 2$ ($K=2$), up to order $|\mathbf{r}| = 3$ ($K=3$), and up to order $|\mathbf{r}| = 4$ ($K=4$), respectively. \\

To determine the best value for $K$, we assessed the performance of the Poisson mixed models estimated using $(\mathbf{y}, \mathbf{X})_{ps_2}$, $(\mathbf{y}, \mathbf{X})_{ps_3}$, and $(\mathbf{y}, \mathbf{X})_{ps_4}$ relative to the true parameters and to the results based on $(\mathbf{y}, \mathbf{X})_d$. We computed bias distributions and 95\% confidence intervals, as well as coverage probabilities. We also compared model selection via AIC and the predictions derived from the selected model for each data type.

\section{Simulation results for Poisson mixed models}\label{section:simres}
\begin{figure}[H]
    \centering
    \includegraphics[width=0.65\linewidth, angle = -90]{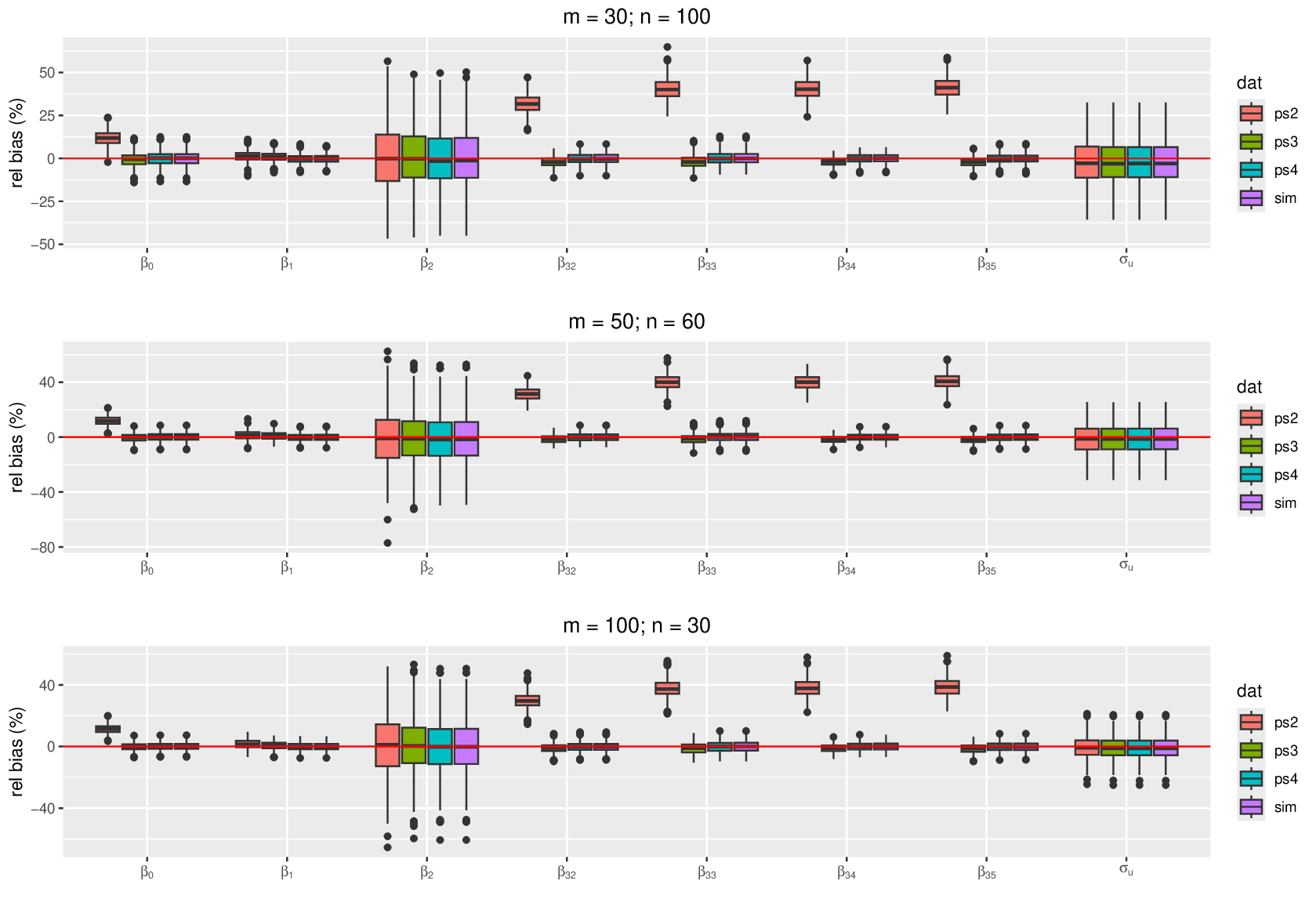}
    \caption{Relative bias distributions of estimates from pseudo-data and simulated data across 500 simulations of a Poisson mixed model under three settings ($m=30$ \& $n=100$; $m=50$ \& $n=60$; and $m=100$ \& $n=30$ where $m$ is the number of groups and $n$ is the uniform group size). For each simulated data, pseudo-data were generated to match sample moments up to (1) $|\mathbf{r}| = 2$ (ps2); (2) $|\mathbf{r}| = 3$ (ps3); and (3) $|\mathbf{r}| = 4$ (ps4).}
    \label{fig:bias}
\end{figure}

\begin{figure}[H]
    \centering
    \includegraphics[width=0.65\linewidth, angle = -90]{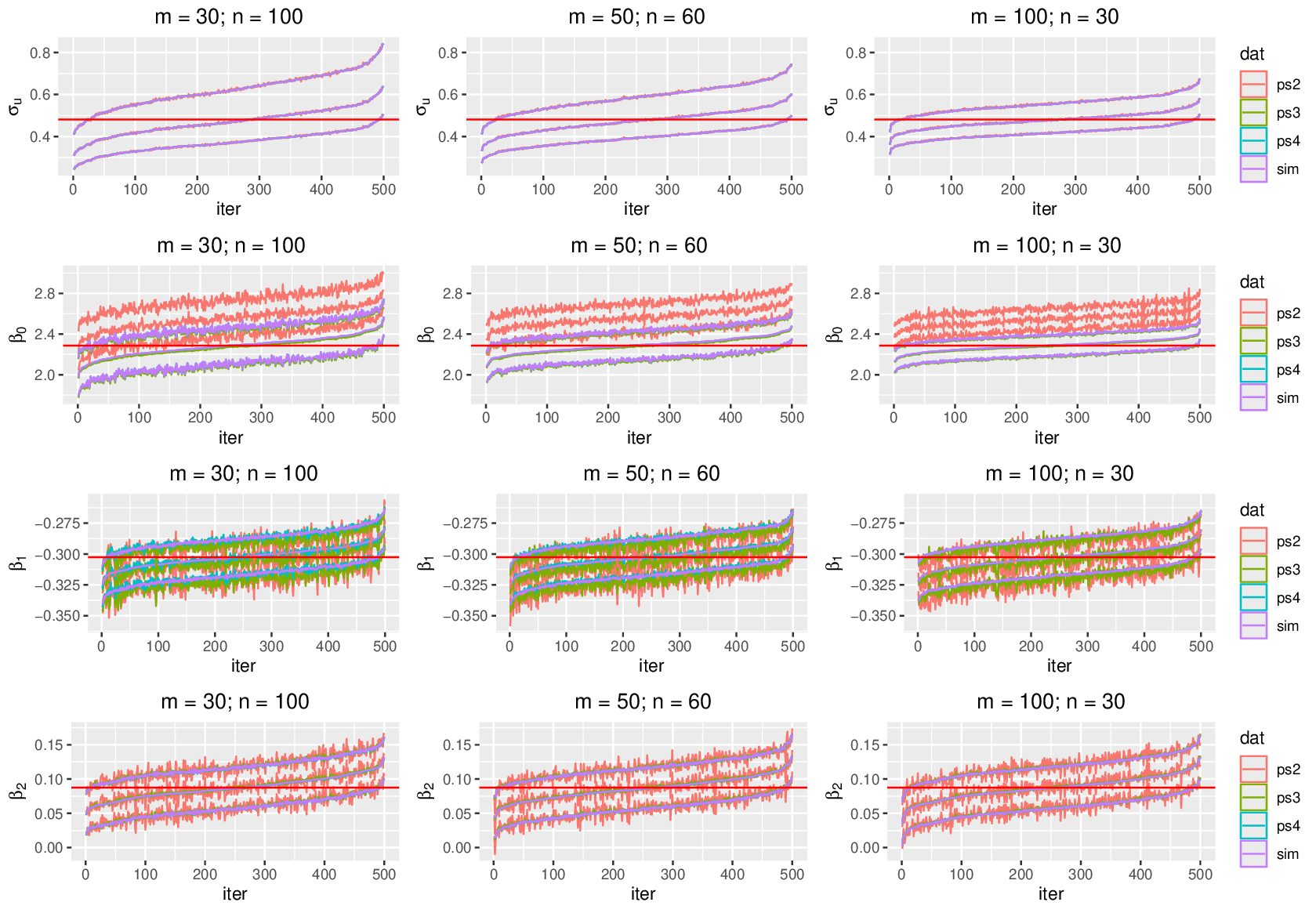}
    \includegraphics[width=0.65\linewidth, angle = -90]{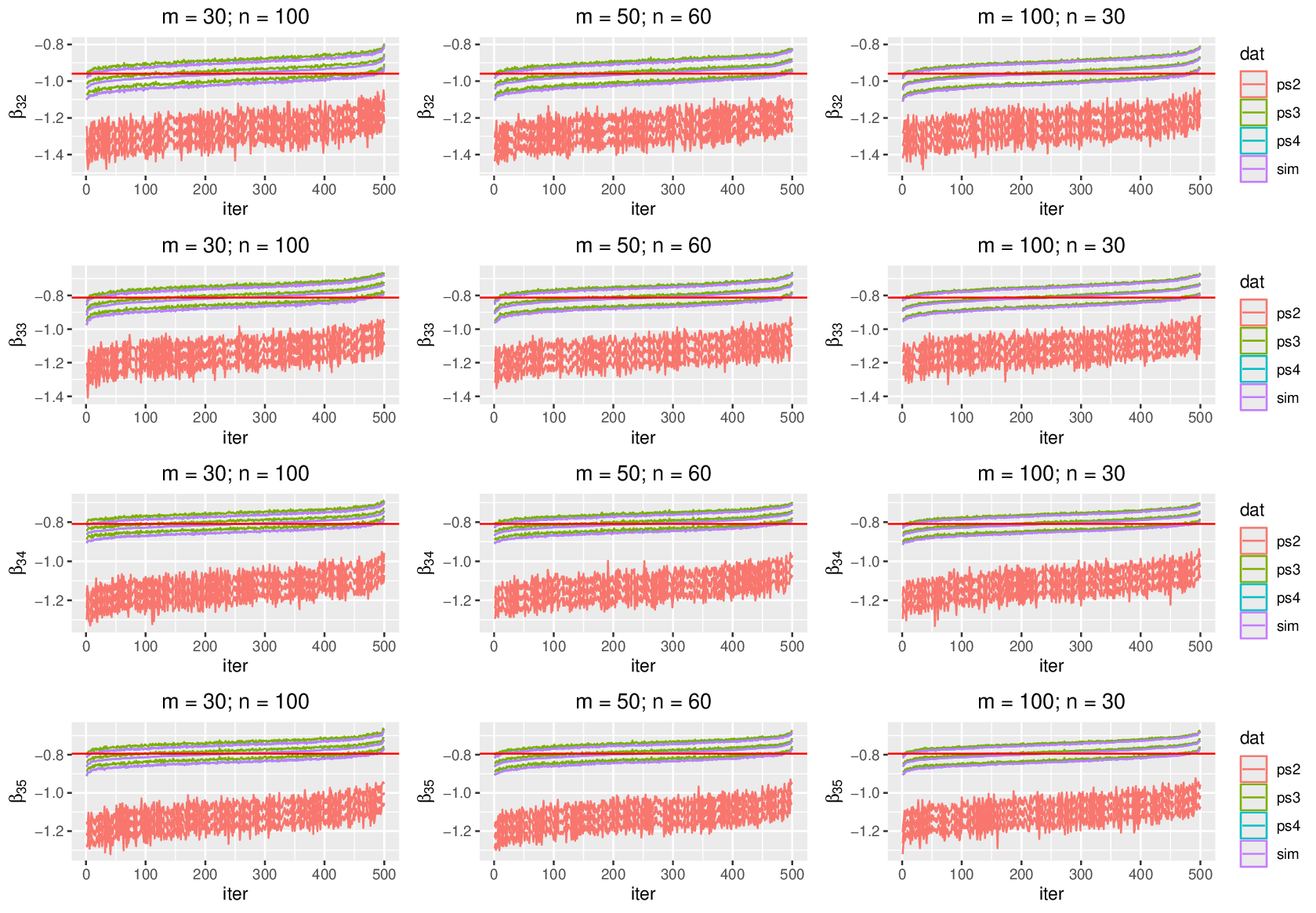}
    \caption{95\% confidence intervals computed on pseudo-data and simulated data across 500 simulations of a Poisson mixed model under three settings ($m=30$ \& $n=100$; $m=50$ \& $n=60$; and $m=100$ \& $n=30$ where $m$ is the number of groups and $n$ is the uniform group size). For each simulated data, pseudo-data were generated to match sample moments up to (1) $|\mathbf{r}| = 2$ (ps2); (2) $|\mathbf{r}| = 3$ (ps3); and (3) $|\mathbf{r}| = 4$ (ps4).}
    \label{fig:ci}
\end{figure}

\begin{figure}[h]
    \centering
    \includegraphics[width=0.65\linewidth, angle = -90]{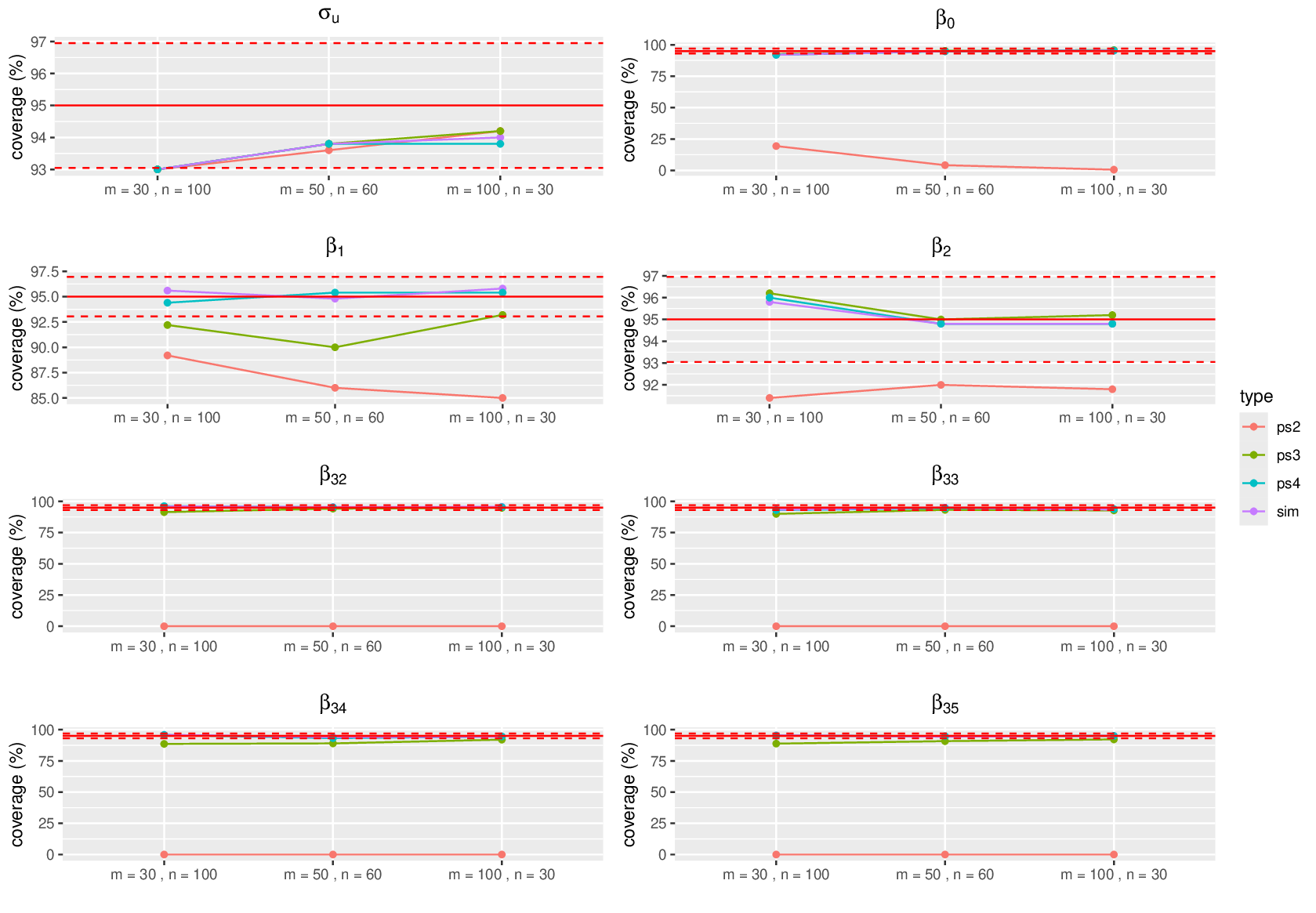}
    \caption{95\% confidence interval coverage computed on pseudo-data and simulated data across 500 simulations of a Poisson mixed model under three settings ($m=30$ \& $n=100$; $m=50$ \& $n=60$; and $m=100$ \& $n=30$ where $m$ is the number of groups and $n$ is the uniform group size). For each simulated data, pseudo-data were generated to match sample moments up to (1) $|\mathbf{r}| = 2$ (ps2); (2) $|\mathbf{r}| = 3$ (ps3); and (3) $|\mathbf{r}| = 4$ (ps4).}
    \label{fig:coverage}
\end{figure}

We first evaluate the estimates produced from $(\mathbf{y}, \mathbf{X})_{ps}$ for different values of $K$. Figure \ref{fig:bias} shows that estimates produced from $(\mathbf{y}, \mathbf{X})_{ps_2}$ performed the poorest in terms of relative bias distribution. Likewise, 95\% confidence intervals from $(\mathbf{y}, \mathbf{X})_{ps_2}$ (Figure \ref{fig:ci}) clearly differ from those produced using $(\mathbf{y}, \mathbf{X})_{d}$. More importantly, these confidence intervals do not capture the true parameter value most of the time, as evidenced by their poor coverage probability (Figure \ref{fig:coverage}). On the contrary, models estimated using $(\mathbf{y}, \mathbf{X})_{ps_4}$ produced the most similar results to models estimated using $(\mathbf{y}, \mathbf{X})_{d}$ in terms of bias and coverage probability. Meanwhile, estimates from $(\mathbf{y}, \mathbf{X})_{ps_3}$ exhibit slight undercoverage for $\beta_1$, $\beta_{32}$, and $\beta_{33}$, and a more pronounced undercoverage for $\beta_{34}$ ($88.6\%$ and $89\%$ for settings 1 \& 2, respectively) and $\beta_{35}$ ($88\%$ for setting 1). Overall, these results indicate that using $(\mathbf{y}, \mathbf{X})_{ps_4}$ yields the best estimates.\\

To assess model selection based on pseudo-data, we estimated all possible models (with linear main effects only) on the simulated and pseudo-data. Since AIC is prone to select more complex models, we added two nuisance variables: one continuous variable ($x_4$) and one binary variable ($x_5$). The correct model includes $x_1$, $x_2$, and $x_3$ only. We compared the AIC values for a total of 31 models consisting of different combinations of the 5 variables. Comparison was made only among models derived from the same dataset; that is, models estimated using simulated data and pseudo-data were not compared to each other. Table \ref{tab:modselAIC} presents the results of calculating the proportion of times the correct model is selected via AIC and the proportion of times the model selected when using the simulated and pseudo-data is the same. \\

The results show that the proportion of times that model selection based on $(\mathbf{y}, \mathbf{X})_d$ leads to the correct model is almost the same as the results for $(\mathbf{y}, \mathbf{X})_{ps_3}$ and $(\mathbf{y}, \mathbf{X})_{ps_4}$ for all combinations of $m$ and $n$. When examining the similarity of model selection results between the simulated and pseudo-data, $(\mathbf{y}, \mathbf{X})_{ps_4}$ selected the same model as $(\mathbf{y}, \mathbf{X})_d$ most of the time. The same can be said for $(\mathbf{y}, \mathbf{X})_{ps_3}$ although the proportion of similarity is slightly lower. The results for $(\mathbf{y}, \mathbf{X})_{ps_2}$ were the most different from $(\mathbf{y}, \mathbf{X})_{d}$ and the proportion of selecting the correct model based on $(\mathbf{y}, \mathbf{X})_{ps_2}$ was the lowest. These findings highly suggest that model selection is identical between using the actual data and $(\mathbf{y}, \mathbf{X})_{ps_4}$, despite truncating the AIC calculation for models based on pseudo-data. As long as model comparison is performed on the same pseudo-data, using the truncated AIC values should not be an obstacle in arriving at the same model as the one that would have been chosen had the actual data been available.\\

\begin{table}[h]
    \centering
    \caption{Proportion of times the selected model via AIC is correct and proportion of times the model selection results are the same between the simulated and pseudo-data (\%)}
    \begin{tabular}{lcccc}
        \hline
         & $(\mathbf{y}, \mathbf{X})_d$ & $(\mathbf{y}, \mathbf{X})_{ps_2}$ & $(\mathbf{y}, \mathbf{X})_{ps_3}$ & $(\mathbf{y}, \mathbf{X})_{ps_4}$ \\
        \hline
        \% correct &  &  \\
        $m=30,n=100$ & 74.5 & 64.0 & 73.5 & 73.5 \\
        $m=50,n=60$ & 67.5 & 64.0 & 66.5 & 67.0 \\
        $m=100,n=30$ & 72.0 & 65.0 & 71.5 & 72.5 \\
        \\
        \% same with $(\mathbf{y}, \mathbf{X})_d$ &  &  \\
        $m=30,n=100$ &  & 76.0 & 97.0 & 99.0 \\
        $m=50,n=60$ &  & 77.5 & 95.5 & 99.5 \\
        $m=100,n=30$ & & 84.0 & 97.5 & 99.5 \\
        \hline
    \end{tabular}
    \label{tab:modselAIC}
\end{table}

\begin{figure}[h]
    \centering
    \includegraphics[width=0.75\linewidth]{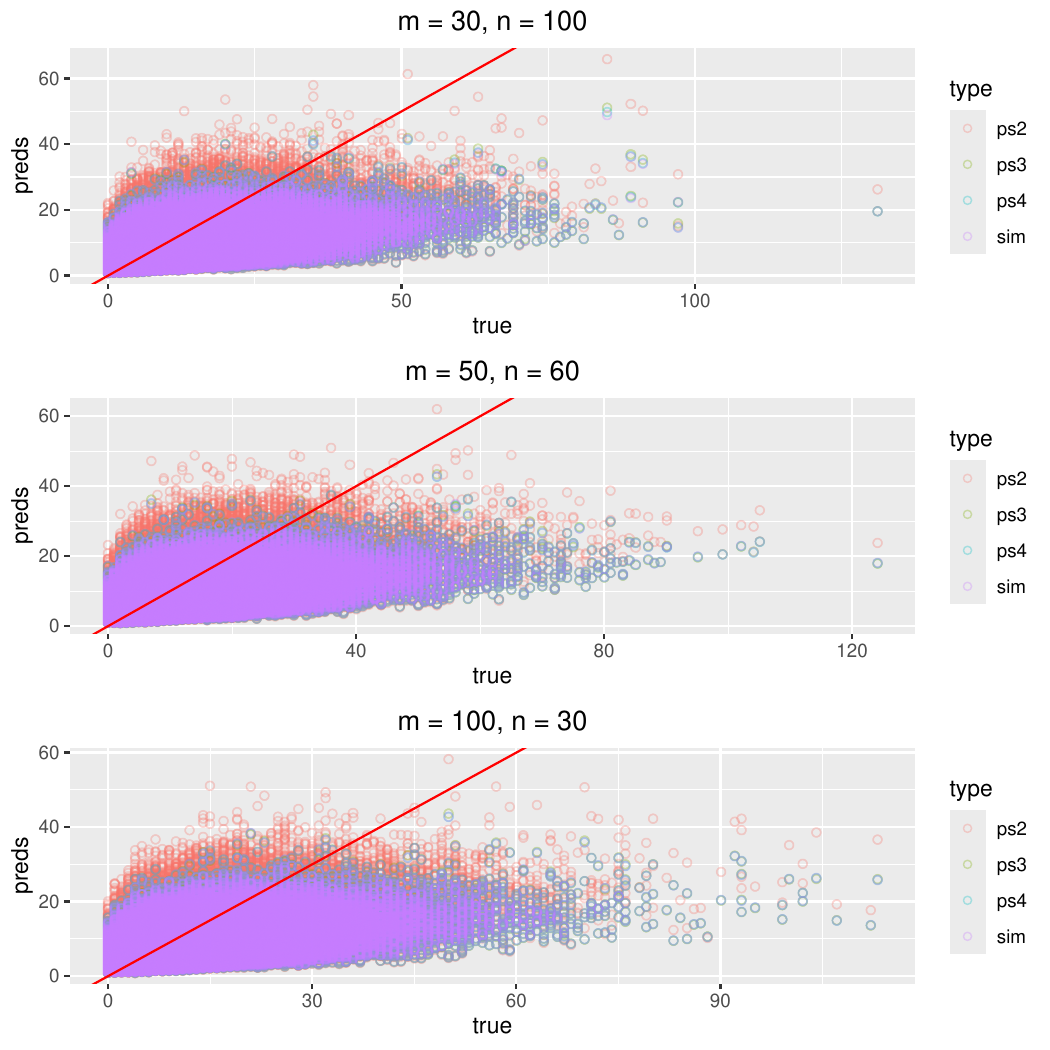}
    \caption{Predictions of models based on pseudo-data and simulated data across 500 simulations of a Poisson mixed model under three settings ($m=30$ \& $n=100$; $m=50$ \& $n=60$; and $m=100$ \& $n=30$ where $m$ is the number of groups and $n$ is the uniform group size). For each simulated data (sim), pseudo-data were generated to match sample moments up to (1) $|\mathbf{r}| = 2$ (ps2); (2) $|\mathbf{r}| = 3$ (ps3); and (3) $|\mathbf{r}| = 4$ (ps4).}
    \label{fig:preds}
\end{figure}

Finally, we examine the predictions of models estimated on pseudo- and simulated data. We used the simulated predictors $\mathbf{X}_{d}$ as input to predict responses. Figure \ref{fig:preds} displays these predictions against the true simulated responses. The results are consistent with previous findings: predictions associated with $(\mathbf{y}, \mathbf{X})_{ps_3}$ and $(\mathbf{y}, \mathbf{X})_{ps_4}$ are quite comparable to the fitted values of models based on $(\mathbf{y}, \mathbf{X})_{d}$. This similarity is exhibited by the overlapping predictions from models estimated using $(\mathbf{y}, \mathbf{X})_d$, $(\mathbf{y}, \mathbf{X})_{ps_3}$, and $(\mathbf{y}, \mathbf{X})_{ps_4}$. On the other hand, predictions from models based on $(\mathbf{y}, \mathbf{X})_{ps_2}$ deviate the most from predictions made using models estimated on $(\mathbf{y}, \mathbf{X})_d$.

\section{Illustrative example using SPARCS 2022}\label{section:realexample}
In this section, we use the findings from the simulation results to demonstrate our proposed framework on the SPARCS 2022 data. The variables included in this demonstration are gender (female or male), length of hospital stay (in days), COVID-19 test result (negative or positive), emergency department indicator (no or yes), and total charges (in USD). Among the 206 hospitals contributing to a total of 2,103,433 patient data, only 205 hospitals with a total of 2,095,246 patient records were actually used after filtering out incomplete cases and invalid data values. The entire framework consists of three steps: (1) computing summary statistics; (2) generating pseudo-data; and (3) estimating models. \\

\subsection{Computing summary statistics}
The first step in the framework and the only task for data providers is data aggregation. Each data provider (i.e., hospital) is expected to compute summary statistics from complete cases. Categorical variables such as gender must be converted to dummy variables before computing summary statistics. These summary statistics are the univariate and multivariate sample moments (Appendix \ref{app:sample_moments}) up to $|\mathbf{r}|=4$, including the number of observations. Summary statistics computed on the standardized variables should also be provided because unscaled values usually cause numerical problems during pseudo-data generation and model estimation. To give an example, Table \ref{tab:summ} displays the univariate summary statistics for A.O. Fox Memorial Hospital while the multivariate sample moments are found in Appendix \ref{app:aohosp}. In this table, the columns \textit{std.} $\bar{\mu}^{(3)}$ and \textit{std.} $\bar{\mu}^{(4)}$ are the univariate third and fourth sample moments, respectively, computed on the standardized variables. In the case of one organization holding all data from all hospitals, summary statistics should be computed per hospital.\\

\begin{table}[h]
    \centering
    \caption{Univariate summary statistics}
    \label{tab:summ}
    \resizebox{\textwidth}{!}{
    \begin{tabular}{lrrrrrrrr}
        \hline
         Variable & Type & $n$ & $\bar{\mathbf{x}}$ & $s^2$ & $\bar{\mu}^{(3)}$ & $\bar{\mu}^{(4)}$ & std. $\bar{\mu}^{(3)}$ & std. $\bar{\mu}^{(4)}$\\
         \hline
         Length of stay (in days) & num & 1520 & 5.15 & 26.45 & 1260.04 & $5.55e^{+04}$ & 5.26 & 51.26 \\
         Total charges (in USD) & num & 1520 & $1.69e^{+04}$ & $1.33e^{+08}$ & $1.51e^{+13}$ & $7.44e^{+17}$ & 2.31 & 11.13\\
         Gender (Male) & bin & 1520 (748) & 0.49 & 0.25 & 0.49 & 0.49 & 0.03 & 1.00\\
         COVID-19 (Positive) & bin & 1520 (183) & 0.12 & 0.11 & 0.12 & 0.12 & 2.33 & 6.43\\
         Emergency (Yes) & bin & 1520 (1013) & 0.67 & 0.22 & 0.67 & 0.67 & -0.71 & 1.50\\
         \hline
         {\footnotesize num - numeric; bin - binary; std. - standardized}
    \end{tabular}
    }
\end{table}

\subsection{Pseudo-data generation}
Once the sample moments up to $|\mathbf{r}|=4$ are available to the data analyst, unconstrained pseudo-data can be generated as outlined in Section \ref{section:psgen}. Table \ref{tab:head_ps_vs_actual} displays the first five observations from both the actual and pseudo-data for the A.O. Fox Memorial Hospital. This table shows that the actual observations are not identical to the pseudo-data, thereby supporting our claim that reconstructing the actual data is improbable through our proposed strategy. This privacy-preserving goal is achieved whenever there are more pseudo-values ($np$) to be generated than the number of sample moments ($\nu$). However, when $np$ is too large, numerical issues begin to arise during the implementation of pseudo-data generation.\\

\begin{table}[h]
    \centering
    \caption{Comparing actual and pseudo-observations for the A.O. Fox Memorial Hospital}
    \label{tab:head_ps_vs_actual}
    \resizebox{\textwidth}{!}{
    \begin{tabular}{lrrrrr}
        \hline
         Type & Length of Hospital & Total Charges & Gender (Male) & COVID19 & Emergency \\
          & Stay (Days) & (USD) &  & (Positive) & (Yes) \\
         \hline
         Actual & 8 & 19,074.29 & 0 & 0 & 1 \\
         & 4 & 13,786.11 & 0 & 0 & 1 \\
         & 6 & 34,246.07 & 1 & 1 & 1 \\
         & 14 & 32,585.40 & 0 & 0 & 1 \\
         & 6 & 12,935.03 & 1 & 0 & 1 \\
         &  &  &  &  &  \\
         Pseudo & 5.096 & 17,021.261 & 0.999 & 1.5e-05 & 0.999 \\
         & 5.706 & 14,191.063 & 0.999 & 2.2e-05 & 0.999 \\
         & 2.536 & 24,598.196 & -6.9e-06 & 2.6e-05 & -2.2e-05 \\
         & 3.847 & 12,557.193 & 0.999 & 8.2e-06 & 0.999 \\
         & 14.685 & 38,028.092 & -3.1e-05 & -5.4e-05 & 1.000 \\
         \hline
    \end{tabular}
    }
\end{table}
When $np$ is too large for some data providers, pseudo-data generation encounters practical limitations. In our proposed strategy, the size of pseudo-data to be generated per group must equate the size of the actual data. For some hospitals in the SPARCS 2022 dataset, sample sizes are extremely huge (e.g. 30,201 observations), overloading the computer's memory during pseudo-data generation. Thus, we propose to generate pseudo-data in subgroups rather than generating all pseudo-observations at once. \\

For the SPARCS 2022 data, we generated pseudo-data for every subgroup of 250 observations for hospitals having more than 500 patient records. This strategy requires each hospital to arbitrarily partition the actual data into subgroups of 250 complete observations. We recommend arbitrary partitioning instead of partitioning based on similarity to avoid having subgroups with few observations, which increases the risk of identifying individuals. If there are subgroups with less than 250 observations, these incomplete subgroups will be merged with one of the complete subgroups. This merging makes the maximum number of observations per subgroup equal to 500. Summary statistics for each hospital will then be composed of summary statistics computed on each subgroup. We set the subgroup size (250) to at least twice the number of target sample moments ($\nu = 125$) to ensure that pseudo-data generation will have a negligible risk of reconstructing the actual data since $np > \nu$. Moreover, a subgroup size between 250 to 500 is still small enough for a regular computer's memory to handle pseudo-data generation without crashing. 

\subsection{Model estimation}
Using the generated pseudo-data, the data analyst may estimate either a linear, logistic, or Poisson regression model with or without random effects. Tables \ref{tab:linear_ps_vs_actual}, \ref{tab:logit_ps_vs_actual}, and \ref{tab:poi_ps_vs_actual} display the linear, logistic, and Poisson mixed models, respectively, estimated on pseudo- and actual data. From the results, we see that the point and interval estimates are very similar when using the pseudo-data in place of the actual data. For the Poisson mixed model's information theoretic measures (AIC and BIC), we also computed the truncated values for the model based on actual data to make them comparable to those of the pseudo-data model.

\begin{table}[h]
\caption{Linear mixed model with Total Charges (USD) as response}
\resizebox{\textwidth}{!}{\color{black}
    \begin{tabular}{lrrrr}
        \hline
        & \multicolumn{2}{c}{pseudo-data} & \multicolumn{2}{c}{actual data} \\
        & Est(std. err.) & 95\% CI (Wald) & Est(std. err.) & 95\% CI (Wald) \\
        \hline
        (Intercept) & -0.198 (0.031)\textcolor{white}{***} & (-0.2594, -0.1371) & -0.198 (0.031)\textcolor{white}{***} & (-0.2594, -0.1371) \\
        Std. Length of Stay (Days) & 0.694 (4.7e-04)\textcolor{white}{***} & (0.6931, 0.6949) & 0.694 (4.7e-04)\textcolor{white}{***} & (0.6931, 0.6949 )\\
        COVID19 (Positive) & -0.039 (0.003)\textcolor{white}{***} & ( -0.0443, -0.0327 ) &-0.039 (0.003)\textcolor{white}{***}& ( -0.0443, -0.0327 )\\
        Gender (Male) & 0.047 (9.3e-04)\textcolor{white}{***} & ( 0.0451, 0.0487 ) &0.047 (9.3e-04)\textcolor{white}{***}& ( 0.0451, 0.0487 )\\
        Emergency (Yes) & -0.083 (0.001)\textcolor{white}{***} & (-0.0853, -0.0812 )& -0.083 (0.001)\textcolor{white}{***}&(-0.0853, -0.0812 ) \\
        $\sigma_{Int}$ & 0.445{\textcolor{white}{***}}&   & 0.445{\textcolor{white}{***}}     &     \\
        \hline
        AIC                           & 4248016    &          & 4248006              \\
        BIC                           & 4248104     &         & 4248094              \\
        total number of patients  & 2,095,246           &         & 2,095,246 \\
        number of hospitals     & 205     &          & 205                       \\
        \hline
    \end{tabular}}
    \label{tab:linear_ps_vs_actual}
\end{table}

\begin{table}[h]
\caption{Logistic mixed model with COVID19 status (Positive or Negative) as response}
\resizebox{\textwidth}{!}{\color{black}
    \begin{tabular}{lrrrr}
        \hline
        & \multicolumn{2}{c}{pseudo-data} & \multicolumn{2}{c}{actual data} \\
        & Est(std. err.) & 95\% CI (Wald) & Est(std. err.) & 95\% CI (Wald) \\
        \hline
        (Intercept) & -5.460 (0.063){***} & (-5.5837, -5.3372) & -5.460 (0.062){***} & (-5.5836, -5.3371) \\
        Std. Length of Stay (Days) & 0.149 (0.005){***} & (0.1385, 0.1597) & 0.149 (0.005){***} & (0.1385, 0.1597 )\\
        Std. Total Charges (USD) & -0.073 (0.007){***} & ( -0.0876, -0.0587 ) &-0.073 (0.007){***}& ( -0.0875, -0.0587 )\\
        Gender (Male) & 0.054 (0.009){***} & ( 0.0369, 0.0719 ) &0.054 (0.009){***}& ( 0.0369, 0.0719 )\\
        Emergency (Yes) & 2.300 (0.020){***} & (2.2602, 2.3402 )& 2.300 (0.020){***}&(2.2600, 2.3400 ) \\
        $\sigma_{Int}$ & 0.828{\textcolor{white}{***}}&   & 0.828{\textcolor{white}{***}}     &     \\
        \hline
        AIC                           & 458967.5    &          & 458969.4              \\
        BIC                           & 459042.8     &         & 459044.8              \\
        total number of patients    & 2,095,246    &    & 2,095,246              \\
        number of hospitals     & 205             &          & 205              \\
        \hline
        \multicolumn{3}{l}{\footnotesize{$^{***}p<0.001$; $^{**}p<0.01$; $^{*}p<0.05$}}
    \end{tabular}}
    \label{tab:logit_ps_vs_actual}
\end{table}

\begin{table}[h]
\caption{Poisson mixed model with Length of Hospital Stay (Days) as response}
\resizebox{\textwidth}{!}{\color{black}
    \begin{tabular}{lrrrr}
        \hline
        & \multicolumn{2}{c}{pseudo-data} & \multicolumn{2}{c}{actual data} \\
        & Est(std. err.) & 95\% CI (Wald) & Est(std. err.) & 95\% CI (Wald) \\
        \hline
        (Intercept) & 1.508 (0.033){***} & (1.4434, 1.5734) & 1.508 (0.033){***} & (1.4434, 1.5734) \\
        COVID19 (Positive) & 0.155 (0.002){***} & (0.1516, 0.1581) & 0.155 (0.002){***} & (0.1516, 0.1581 )\\
        Std. Total Charges (USD) & 0.130 (4.9e-05){***} & ( 0.1295, 0.1297 ) &0.130 (4.9e-05){***}& ( 0.1295, 0.1296 )\\
        Gender (Male) & 0.094 (5.8e-04){***} & ( 0.0928, 0.0951 ) &0.094 (5.8e-04){***}& ( 0.0928, 0.0951 )\\
        Emergency (Yes) & 0.323 (6.9e-04){***} & (0.3219, 0.3246 )& 0.323 (6.9e-04){***}&(0.3219, 0.3246 ) \\
        $\sigma_{Int}$ & 0.473{\textcolor{white}{***}}&   & 0.473{\textcolor{white}{***}}     &     \\
        \hline
        (truncated) AIC                           & (-21244749)    &          & (-21244336) 16242582              \\
        (truncated) BIC                           & (-21244674)     &         & (-21244261) 16242658              \\
        total number of patients                     & 2,095,246           &         & 2,095,246                    \\
        number of hospitals     & 205             &          & 205                       \\
        \hline
        \multicolumn{3}{l}{\footnotesize{$^{***}p<0.001$; $^{**}p<0.01$; $^{*}p<0.05$}}
    \end{tabular}}
    \label{tab:poi_ps_vs_actual}
\end{table}


\section{Discussion}\label{section:discussion}
In this paper, we proposed a framework that enables GLM[M] estimation without requiring access to individual-level data. This framework requires data providers to share only the sample moments of the actual data in one round of communication. We have shown that these sample moments must include all univariate and multivariate moments up to the 4th order. To facilitate model estimation through existing functionalities in statistical software programs like \texttt{R}, we proposed to generate and use pseudo-data rather than directly using sample moments for model fitting. These pseudo-data are generated such that their sample moments match those of the actual data. Our simulations and illustrative example involving real-world data confirm the viability of using pseudo-data as a proxy for the actual data during model estimation.

Expressing the log-likelihood as a polynomial plus an error term plays a crucial role in why and how our approach works. Through the Weierstrass approximation theorem, a polynomial expression of any continuous function, such as logarithmic and exponential functions, is always possible within an error $\epsilon > 0$. When the terms of the log-likelihood are written as polynomials with respect to the data ($\mathbf{y}$ and $\mathbf{X}$), the summation is easy to distribute over the terms. This distributive property makes summary statistics more explicit. Thus, any log-likelihood can always be expressed in terms of summary statistics, which in turn can be computed from sample moments. It follows that the log-likelihood can be constructed up to some level of accuracy using sample moments and does not necessitate access to the actual individual-level data. However, instead of directly using sample moments, our framework proposes to generate pseudo-data. \\

The purpose of pseudo-data goes deeper than the convenience of using existing functionalities for model estimation. Directly using sample moments to construct the log-likelihood implies using a polynomial to approximate the log-likelihood. In this polynomial-approximated log-likelihood, the error term is assumed to be zero. This assumption may be too strict because of the new structure imposed on the log-likelihood. ML estimation on this polynomial-approximated log-likelihood may lead to biased parameter estimates. An even worse possibility is the failure to converge to an optimum solution. With pseudo-data, the structure of the log-likelihood is retained and the error term is not truncated. Log-likelihood constructed from pseudo-data resembles the log-likelihood from actual data more closely than the polynomial-approximated log-likelihood. As a consequence, ML estimates are more similar. \\

The similarity in performance of log-likelihoods constructed from actual and pseudo-data can be made as close as possible only through the polynomial part. The polynomial part of the re-expression of the log-likelihood is composed of summary statistics which can be computed from sample moments. When the sample moments between the actual and pseudo-data are nearly identical, the polynomial part of the log-likelihood is nearly identical too. However, if the error between the polynomial and the log-likelihood is large, closely matched sample moments become useless. This error cannot be fully controlled but can only be minimized by using higher degrees $K$ in the polynomial part of the log-likelihood. \\

The simulations indicate that setting $K=4$ produces the best parameter estimates for Poisson mixed models under our proposed framework, relative to the other values that were investigated ($K=2,3$). Equivalently, the sample moments of pseudo-data must match those of the actual data up to $|\mathbf{r}|=4$. This result is consistent with the implicit goal of minimizing the error part in the polynomial expression of the log-likelihood. Following this reasoning, considering values $K > 4$ might be tempting. \\

While considering even higher polynomial terms is mathematically possible, it becomes impractical. When the size of target sample moments $\nu$ becomes too large, generating pseudo-data for small-sized groups becomes problematic. It may either reconstruct the actual observations or lead to large discrepancies in sample moments between the actual and pseudo-data. The discrepancies arise whenever pseudo-data are too constrained because the number of pseudo-data values is not large enough to allow flexibility when satisfying such constraints. These discrepancies affect the similarity between log-likelihoods based on actual and pseudo-data, ultimately influencing the outcome of ML estimation. \\ 

ML estimation is also influenced by constraints imposed on the nature of pseudo-values generated. Since our proposed approach lifts the binary and integer constraints, generating pseudo-data that satisfy the sample moments specified by the actual data becomes easier. With a closer match in sample moments between actual and pseudo-data, ML estimation produces estimates that are identical up to the third decimal place, as exhibited in Section \ref{section:realexample}. Moreover, reconstructing the actual observations and identifying characteristics of individuals from the actual data become much less likely. For instance, binary values such as gender for an individual in the actual data may be unrecognizable from a pseudo-data value, which can be any continuous value. In contrast, synthetic data generation retains the nature of the actual variable. \\

While closely related to synthetic data generation, our proposed approach has distinct characteristics. A major edge of our proposed approach is that it only requires the sample moments of the actual data, whereas synthetic data generation needs access to the actual individual-level data \cite{electronics13173509}. In a federated data setting, only data providers have access to individual-level data. If synthetic data are to be generated, each data provider must have the technical capability. On the contrary, our framework allows either data providers or data analysts to generate pseudo-data. In addition, our proposed framework does not assume any distribution on the actual data as well as on the pseudo-data, unlike some statistical-based methods for synthetic data generation \cite{PEZOULAS20242892}. Lastly, pseudo-data generation does not involve hyperparameter tuning, which is the case for machine learning-based synthetic data generation methods \cite{10.3389/frai.2022.918813}. \\ 

Our proposed framework has considerable merits over its predecessors, which handle Gaussian \citep{https://doi.org/10.1002/sim.10300} and Bernoulli \citep{https://doi.org/10.1002/bimj.70080} models only. First, it is generalizable to any exponential family member. Second, it potentially allows estimation of other specialized models from summary statistics, such as geostatistical models. Because of its likelihood-based nature, our proposed approach may even be applicable for Bayesian models. This possibility comes from the versatility of utilizing the generated pseudo-data for various model estimation tasks. However, a non-trivial drawback, as with synthetic data generation methods, is scalability \citep{app14145975}. 

For large scale problems, simultaneous pseudo-data generation for all variables becomes more computationally demanding. This issue arises due to the Jacobian matrix which now has $\nu$ rows and $np$ columns, instead of just $n$ columns at a time. More importantly, the Cholesky decomposition involved in updating the parameter estimates has to deal with $np \times np$ matrices. Consequently, larger sample sizes with more variables overloads the memory of a computer. This issue was indeed encountered when implementing the proposed method on the SPARCS 2022 data. To circumvent this limitation, we proposed partitioning the actual data into arbitrary subgroups within each hospital. In effect, pseudo-data generation does not have to deal with huge matrices.

\section{Conclusion}\label{section:concl}
\noindent Overall, we have shown in this paper that we can generate a compressed version of pseudo-data while matching all the sample moments up to 4th order of the actual data through column generation. These compressed pseudo-data can be used to estimate generalized linear mixed models (GLMM) in a federated data setting involving large datasets. The proposed strategy enables data analysts to implement pseudo-data generation for an originally large-sized actual data without exceeding the memory of a local workstation with limited memory. Consequently, statistical modeling of federated data is neither impeded by privacy restrictions nor the large size of the actual data.

\section{Conflict of Interest}
The author(s) declare(s) no potential conflicts of interest with respect to the research, authorship and/or publication of this article.


\section{Funding}

    This study was supported by the Special Research Fund of Hasselt University (BOF08M01, Methusalem grant).

\bibliographystyle{smama.bst}
\bibliography{ref.bib}

\appendix
\section{Appendix}
\subsection{Theorems}\label{app:theorems}
\begin{theorem}[Weierstrass approximation theorem]
    Suppose $f$ is a continuous real-valued function defined on the real interval $[a,b]$. For every $\varepsilon > 0$, there exists a polynomial $p$ of degree $K$ such that for all $x$ in $[a,b]$, we have $|f(x)-p(x)| < \varepsilon$.
\end{theorem}

\begin{theorem}[Taylor's theorem]
    Suppose $f$ is defined on some open interval $I$ around a real number $\text{a}$ and is at least $K+1$-times differentiable on this interval. Then for each $x \neq a$ in~$I$, there exists a value $\xi$ between $x$ and $a$ such that
    \begin{align}
        f(x) &= \sum_{k=0}^K \frac{f^{(k)}(a)}{k!}\left(x-a\right)^k + \frac{f^{(K+1)}(\xi)}{(K+1)!}\left(x-a\right)^{K+1}\textcolor{blue}{,}
    \end{align}
    where $f^{(k)}$ is the $k$th derivative of the function $f$.
\end{theorem}

\begin{theorem}[L'H$\hat{\text{o}}$pital's rule]
Given functions f and g which are defined on an open interval I and differentiable on the same interval but not necessarily at a constant c, if $\lim_{x\to c} f(x) = \lim_{x\to c} g(x) = 0$ or $\pm \infty$, $g'(x) \neq 0$ for all $x$ in I but not necessarily at c, and $\lim_{x \to c} \frac{f'(x)}{g'(x)}$, then
\begin{align}
    \lim_{x \to c} \frac{f(x)}{g(x)} &= \lim_{x \to c} \frac{f'(x)}{g'(x)}.
\end{align}
    
\end{theorem}


\subsection{PASS derivation for the exponential family}\label{app:poly_loglik}
Consider the log-likelihood of the parameters $\{\boldsymbol{\theta}, \phi\}$ of any exponential family member
\begin{align}\label{eq:loglik}
    l(\boldsymbol{\theta}, \phi) &= \log{\prod_{i=1}^n \exp{\left[\frac{y_i\theta_i - b(\theta_i)}{a(\phi)} + c(y_i, \phi)\right]}},
\end{align}
where $\boldsymbol{y}$ is the vector of data with size $n$ and $a(.)$, $b(.)$, and $c(.)$ are known functions that further characterize a member of this family. Re-expressing Equation \ref{eq:loglik} leads to:
\begin{align}
    l(\boldsymbol{\theta}, \phi) &= \sum_{i=1}^n\left[\frac{y_i\theta_i - b(\theta_i)}{a(\phi)} + c(y_i, \phi)\right] \\
    &= \frac{1}{a(\phi)}\left[ \sum_{i=1}^n y_i\theta_i - \sum_{i=1}^n b(\theta_i)\right] + \sum_{i=1}^n c(y_i, \phi).
\end{align}
For members of this family with $\theta = \mathbf{X}\boldsymbol{\beta}$ such as the Gaussian, Bernoulli, and Poisson models, $\sum_{i=1}^n y_i\theta_i$ and $\sum_{i=1}^n c(y_i, \phi)$ can be exactly constructed from aggregate statistics without knowing the individual values $y_i$ and $\mathbf{x}_i$. For instance, it is enough to know the sample size $n$, sample mean ($\bar{\mathbf{y}}$) and variance ($s^2_{\mathbf{y}}$) of the responses, sample covariances ($s_{\mathbf{yx}_j}$) between $\mathbf{y}$ and each of the $p$ predictors $\mathbf{x}_j$, $j=1,...,p$, and $\sum{\log{(\mathbf{y}!)}}$. For $\sum_{i=1}^n b(\theta_i)$ however, it is not as straightforward especially for the Bernoulli and Poisson models. To get around this challenge, we use the Weierstrass approximation theorem (Appendix \ref{app:theorems}) and express $b(\theta_i)$ as

\begin{align}
    b(\theta_i) &= \sum_{k=0}^K{\alpha_k\theta_i^k} + \epsilon_i,
\end{align}

which, for exponential family members with $\theta_i = \mathbf{x}_i^T\boldsymbol{\beta}$, translates to

\begin{align}
    b(\mathbf{x}_i^T \boldsymbol{\beta}) &= \sum_{k=0}^K{\alpha_k(\mathbf{x}_i^T \boldsymbol{\beta})^k} + \epsilon_i.
\end{align}

Here, $K$ is the polynomial degree, $\alpha_k$ is the associated coefficient of the $k$th term, and $\epsilon_i$ is the polynomial approximation error. With this, we can now express $\sum_{i=1}^n b(\theta_i)$ as
\begin{align}
    \sum_{i=1}^n b(\mathbf{x}_i^T \boldsymbol{\beta}) &= \sum_{k=0}^K{\alpha_k \sum_{i=1}^n(\mathbf{x}_i^T \boldsymbol{\beta})^k} + \sum_{i=1}^n \epsilon_i,
\end{align}
where for Gaussian models, $K=2$, $\alpha_0 = \alpha_1 = 0$, $\alpha_2 = 1/2$, and $\epsilon_i = 0$. Further algebraic derivations lead to using sample moments (Appendix \ref{app:sample_moments}) as sufficient statistics to construct this polynomial expression of the log-likelihood up to an error $\epsilon$.   
\subsection{Sample moments}\label{app:sample_moments}
The univariate sample moment of order $r$ of a variable $X$ is defined as 
\begin{align}\label{mu_r}
    \bar{\mu}^{(r)} &= \frac{1}{n}\sum_{i=1}^n{x_i^r}, \hspace{5mm} r = 1,2,...
\end{align}

The bivariate or joint sample moment of order $(r_1,r_2)$ between variables $X_1$ and $X_2$ is
\begin{align}
    \bar{\mu}^{(r_1,r_2)} &= \frac{1}{n}\sum_{i=1}^n{x_{1i}^{r_1}x_{2i}^{r_2}}, \hspace{5mm} r_1, r_2 = 1,2,...
\end{align}

In general, the $p$-variate sample moment of order $\mathbf{r} = (r_1,r_2,...,r_p)$, using a multi-index notation, is
\begin{align}
    \bar{\mu}^{(\mathbf{r})} &= \frac{1}{n}\sum_{i=1}^n {x_{i}^{\mathbf{r}}},
\end{align}

where $|\mathbf{r}| = r_1 + r_2 + ... + r_p$ and $x^{\mathbf{r}} = x_1^{r_1} x_2^{r_2}...x_p^{r_p}$.

\subsection{Log-likelihood of saturated models}\label{app:loglik_sat}

The $i$th log-likelihood contribution in a saturated model for a Bernoulli 
\begin{align}
    y_i \log{(y_i)} + (1-y_i)\log{(1-y_i)},
\end{align}
or Poisson model
\begin{align}
    y_i\log{(y_i)} - y_i - \log{(y!)}
\end{align}

involves taking the logarithm of $y_i$. Whenever $y_i=0$, the term
\begin{align}
    y_i\log{(y_i)} &= 0\log{(0)}
\end{align}
becomes mathematically undefined. However, the limit from the right can be examined as
\begin{align}
    \lim_{y_i\to0^+} y_i\log{(y_i)} &= \lim_{y_i\to0^+} \frac{\log{(y_i)}}{1/y_i} \\
    &= \frac{\lim_{y_i\to0^+} \log{(y_i)}}{\lim_{y_i\to0^+} 1/y_i}.
\end{align}
Since $\lim_{y_i\to0^+} \log{(y_i)} = -\infty$ and $\lim_{y_i\to0^+} 1/y_i = +\infty$, we can use L'H$\hat{\text{o}}$pital's rule (Appendix \ref{app:theorems}) to find that the limit is
\begin{align}
    \lim_{y_i\to0^+} y_i\log{(y_i)} &= \lim_{y_i\to0^+} \frac{1/y_i}{-1/y_i^2} \\
    &= \lim_{y_i\to0^+} -y_i \\
    &= 0.
\end{align}
From this, we can consider $y_i\log{(y_i)}$ to be almost 0. Thus, when $y_i = 0$ or 1 for the Bernoulli case and $y_i=0$ for the Poisson case, the contribution of the log-likelihood of the saturated model to the deviance is almost 0.

\subsection{Multivariate summary statistics for A.O. Fox Memorial Hospital}\label{app:aohosp}

\begin{table}[h]
    \centering
    \caption{Covariance matrix}
    \resizebox{\textwidth}{!}{
    \begin{tabular}{llllll}
        \hline
         \hspace{2cm} Variable 2 & Length of & Total charges & Gender & COVID-19 & Emergency\\
          Variable 1& stay (days) & (USD) & (Male) & (Positive) & (Yes)\\
         \hline
         Length of stay (days) & 1.00 & --- & --- & --- & ---\\
         Total charges (USD) & 0.73 & 1.00 & --- & --- & ---\\
         Gender (Male) & 0.03 & 0.05 & 1.00 & --- & ---\\
         COVID-19 (Positive) & 0.16 & 0.37 & 0.05 & 1.00 & ---\\
         Emergency (Yes) & 0.09 & 0.12 & -0.03 & 0.06 & 1.00\\
         \hline
    \end{tabular}
    }
\end{table}

\begin{table}[h]
    \centering
    \caption{Bivariate sample moments ($|\mathbf{r}| \geq 3$)}
    \begin{tabular}{lllllll}
        \hline
         Variable 1 & Variable 2 & $\bar{\mu}^{(2,1)}$ & $\bar{\mu}^{(1,2)}$ & $\bar{\mu}^{(2,2)}$ & $\bar{\mu}^{(3,1)}$ & $\bar{\mu}^{(1,3)}$\\
         \hline
         Length of & Total charges & 2.87 & 2.04 & 14.32 & 24.5 & 10.6\\
         stay (days) & (USD) \\
         \\
         & Gender & 0.12 & 0.001 & 1.00 & 0.65 & 0.03\\
         & (Male) \\ 
         \\
         & COVID-19 & 0.23 & 0.38 & 1.53 & 0.19 & 1.04\\
         & (Positive) \\
         \\
         & Emergency & 0.20 & -0.07 & 0.86 & 1.65 & 0.14\\
         & (Yes) \\
         \\
         Total charges & Gender & 0.14 & 0.001 & 1.00 & 0.65 & 0.05\\
         (USD) & (Male) & \\
         \\
         & COVID-19 & 0.79 & 0.87 & 2.85 & 3.44 & 2.41\\
         & (Positive) \\
         \\
         & Emergency & 0.15 & -0.08 & 0.89 & 1.13 & 0.18\\
         & (Yes) \\
         \\
         Gender & COVID-19 & 0.002 & 0.12 & 1.00 & 0.05 & 0.34\\
         (Male) & (Positive) \\
         \\
         & Emergency & -0.001 & 0.02 & 1.00 & -0.03 & -0.05\\
         & (Yes) \\
         \\
         COVID-19 & Emergency & 0.15 & -0.05 & 0.89 & 0.41 & 0.10\\
         (Positive) & (Yes) \\
         \hline
    \end{tabular}
\end{table}

\begin{table}[h]
    \centering
    \caption{Trivariate sample moments}
    \begin{tabular}{lllllll}
        \hline
         Variable 1 & Variable 2 & Variable 3 & $\bar{\mu}^{(1,1,1)}$ & $\bar{\mu}^{(2,1,1)}$ & $\bar{\mu}^{(1,2,1)}$ & $\bar{\mu}^{(1,1,2)}$\\
         \hline
         Length of & Total charges & Gender & 0.12 & 0.75 & 0.60 & 0.73\\
         stay (days) & (USD) & (Male) \\
         \\
         & & COVID-19 & 0.39 & 1.20 & 1.96 & 1.65\\
         & & (Positive) \\
         \\
         & & Emergency & 0.23 & 1.29 & 1.16 & 0.57\\
         & & (Yes) \\
         \\
         & Gender & COVID-19 & 0.01 & 0.09 & 0.16 & 0.06\\
         & (Male) & (Positive) \\
         \\
         & & Emergency & 0.01 & 0.05 & 0.09 & 0.02\\
         & & (Yes) \\
         \\
         & COVID-19 & Emergency & 0.03 & 0.16 & 0.17 & 0.14\\
         & (Positive) & (Yes) \\
         \\
         Total charges & Gender & COVID-19 & 0.06 & 0.26 & 0.38 & 0.20\\
         (USD) & (Male) & (Positive) \\
         \\
         & & Emergency & 0.04 & 0.10 & 0.12 & 0.02\\
         & & (Yes) \\
         \\
         & COVID-19 & Emergency & 0.06 & 0.28 & 0.25 & 0.33\\
         & (Positive) & (Yes) \\
         \\
         Gender & COVID-19 & Emergency & 0.04 & 0.07 & 0.05 & 0.03\\
         (Male) & (Positive) & (Yes) \\
         \hline
    \end{tabular}
\end{table}

\begin{table}[h]
    \centering
    \caption{Quadvariate sample moments}
    \begin{tabular}{lllll}
        \hline
         Variable 1 & Variable 2 & Variable 3 & Variable 4 & $\bar{\mu}^{(1,1,1,1)}$\\
         \hline
         Length of & Total charges & Gender & COVID-19 & 0.12\\
         stay (days) & (USD) & (Male) & (Positive)\\
         \\
         & & & Emergency & 0.07\\
         & & & (Yes) \\
         \\
         Length of & Total charges & COVID-19 & Emergency & 0.19\\
         stay (days) & (USD) & (Positive) & (Yes)\\
         \\
         Length of & Gender & COVID-19 & Emergency & 0.001\\
         stay (days) & (Male) & (Positive) & (Yes)\\
         \\
         Total charges & Gender & COVID-19 & Emergency & 0.03\\
         (USD) & (Male) & (Positive) & (Yes)\\
         \hline
    \end{tabular}
\end{table}

\end{document}